\documentstyle[12pt]{article}
\begin{document}

\title{Some aspects of the critical behavior of the \\
Two-Neighbor Stochastic Cellular Automata}

\author{A.Yu. Tretyakov\\
School of Information Sciences, Tohoku University,\\
Aramaki, Aoba-ku, Sendai, 980-77, JAPAN \vspace{3mm}\\
N. Inui\\
Himeji Institute of Technology,\\
2167, Shosha, Himeji, 671-22, JAPAN \vspace{3mm}\\
M. Katori and H. Tsukahara\\
Department of Physics, Faculty of Science and Engineering, Chuo University,\\
Kasuga, Bunkyo-ku, Tokyo 112, JAPAN\\
}

\maketitle

\setlength{\baselineskip}{9mm}
\begin{abstract}
\setlength{\baselineskip}{7mm}

Using Pade approximations and Monte Carlo simulations, we study the phase
diagram of the Two-Neighbor Stochastic Cellular Automata, which have two
parameters $p_{1}$ and $p_{2}$ and include the mixed site-bond directed
percolation (DP) as a special case.
The phase transition line $p_{1}=p_{1 {\rm c}}(p_{2})$ has endpoints
at $(p_{1}, p_{2})=(1/2,1)$ and at (0.8092, 0).
The former point (1/2,1) is a special point at which Compact DP transition
occurs and its critical exponents are known exactly.
Results of time-dependent simulation show that in the whole range of
parameters, excluding this point (1/2,1), the system belongs to the DP
universality class. It is first shown that the shape of the phase transition
line near this special point has, asymptotically, a parabolic shape,
{\it i.e.}, $p_{1{\rm c}}(p_{2})-1/2 \sim (1-p_{2})^{\theta}$
with $\theta = 1/2$ for $0 < 1-p_{2} \ll 1$.
We use the Monte Carlo data to assess the
accuracy of rigorous bounds for the line recently reported by
Liggett and by Katori and Tsukahara. It is also shown that outside the
vicinity of the special point (1/2,1), the curve is well approximated by an
interpolation formula, similar to the one proposed by Yanuka and Englman.

\end{abstract}

PACs number(s): 05.70.Ln,64.90.+b

\section{Introduction}

The two-neighbor stochastic cellular automata (SCA),
introduced by Domany and Kinzel, can be
regarded as a very general model of a spread of influence in 1+1
dimensions. A site on layer $i+1$ is occupied with probability $p_1$ if
one and only one of its nearest neighbors on the previous layer is
occupied, it is occupied with probability $p_2$ if both of the
neighbors are occupied, and it remains vacant if both neighbors are
vacant (Fig.  \ref{fig:model}). Often the system is regarded as a
process taking place in one-dimension, with the layer number $i$
replaced by discrete time $t$.

As one can easily see, the case of $p_1=p_2=p$ is equivalent to the
site directed percolation (DP) with site occupation probability $p$. It is
easy to show that $(p_1=p,p_2=2p-p^2)$ corresponds to bond DP
with the fraction of open bonds $p$. The mixed site-bond
DP with the fraction of open sites $\alpha$ and the
fraction of open bonds $\beta$ corresponds to the two-neighbor SCA
with $p_1=\alpha \beta$ and $p_2=\alpha (2 \beta-\beta^2)$
\cite{kinz,kinz2,durr}.

In the present work, by using Monte Carlo simulations, we evaluate the
phase transition line which is characterized as follows.
We consider the process starting from a single occupied particle
and let $P(t)$ be the probability that at least one particle survives
at time $t$. In the subcritical phase $p_{1} < p_{1{\rm c}}(p_{2})$,
this survival probability decreases exponentially to zero, while in
the supercritical phase $p_{1} > p_{1{\rm c}}(p_{2})$, it converges to
a positive value: $P_{\infty} \equiv \lim_{t \rightarrow \infty} P(t) >0$.
On the phase transition line $p_{1}=p_{1{\rm c}}(p_{2})$,
$P(t)$ shows a power-law decay and critical phenomena are observed.
The ultimate survival probability in the supercritical phase
behaves as follows in the vicinity of the phase transition line:
$P_{\infty}(p_{1},p_{2}) \sim (p_{1}-p_{1 {\rm c}}(p_{2}))^{\beta}$
as $p_{1} \rightarrow p_{1 {\rm c}}(p_{2})$ for each $p_{2}$.
It is conjectured for $p_{2} < 1$ that the critical
exponent $\beta$ is universal,{\it i.e.}, independent of $p_{2}$
and is the same as the pure DP, $\beta \simeq 0.276$.

The two-neighbor SCA with $p_2=1$ is equivalent to the
reaction-limited case of the so-called dimer-dimer model \cite{takin},
this case (sometimes referred to as Compact Directed Percolation) is
understood rather well \cite{kinz,durr,dick}.
In particular, it is easy to see
$P_{\infty}(p_{1},1)= 1-(1-p_{1})^{2}/p_{1}^{2} \simeq 8(p_{1}-1/2)$
for $p_{1} \simeq 1/2$, which means $p_{1 {\rm c}}=1/2 $ and
$\beta(p_{2}=1)=1$ \cite{durr}.

The point $(p_{1},p_{2})=(1/2,1)$ is an endpoint of the critical line
$p_{1}=p_{1 {\rm c}}(p_{2})$ in the phase diagram of the two-neighbor
SCA. The universality for $p_{2} <1$ and the above mentioned exact
result for $p_{2}=1$ imply that we will observe the following phenomenon
called a crossover when $p_{2} \simeq 1$ \cite{durr}.
There is a function $\varepsilon(p_{2}) \geq 0 $ such that if
$0 < p_{1}-p_{1 {\rm c}}(p_{2}) \ll \varepsilon(p_{2})$ then
$P_{\infty} \sim (p_{1} - p_{1 {\rm c}}(p_{2}) )^{\beta}$ with
$\beta \simeq 0.276$, while if
$\varepsilon(p_{2}) < p_{1} - p_{1 {\rm c}}(p_{2}) \ll 1$ then
$P_{\infty} \sim (p_{1}-1/2)$ and that $\varepsilon(p_{2}) \rightarrow 0$
as $p_{2} \rightarrow 1$.

In order to study the universality and the crossover, it is necessary to
make precise evaluation of the phase transition line.
In particular, we need to know the asymptotic behavior of the
phase transition line near the special point $(p_{1},p_{2})=(1/2,1)$.
By using series expansion technique and
comparing the results with the data obtained by Monte
Carlo we conclude that the asymptotic shape of the phase transition line
near $(1/2, 1)$ is parabolic. In other words,
$p_{1 {\rm c}}(p_{2}) -1/2 \sim (1-p_{2})^{\theta}$ with $\theta=1/2$
for $0 < 1-p_{2} \ll 1$. We use the Monte Carlo data to assess the
accuracy of rigorous bounds recently obtained theoretically \cite{kattu,ligg}.

\section{Monte Carlo simulations}

Two points of the phase diagram, corresponding to pure site
and bond DP, can be estimated from the
corresponding threshold values as  (0.705485,0.705485) and
(0.644701,0.873762), using the data taken from \cite{ndic}. The sector
of the phase diagram between them corresponds to the mixed site-bond
DP and can be obtained using the data given in
\cite{mixed}.  Therefore, to obtain a complete phase diagram only areas
with $0.5<p_1<0.644701$ and $p_1>0.705485$ need to be covered.

Similarly to \cite{mixed}, to determine the location of the
phase-transition line, we used the time dependent simulation technique
and analyzed the time dependence of the local slopes for the mean
number of particles $n$,  for the survival probability $P$ and for the mean
square radius $R$.  Simulations were conducted starting from a single
occupied particle, taking 1000 steps and averaging over $10^5$
configurations.  Some of the local slopes curves (for the mean number
of particles) are given in Fig. \ref{fig:slopes}. Our
results agree with the assumption that the two-neighbor SCA
belong to the DP universality class for $p_2<1$
\cite{dick}, with the dynamical critical exponents defined as

\begin{equation}
\begin{array}{ccc}
P(t) & \propto &t^{-\delta} \\
n(t) & \propto &t^{\eta} \\
R^{2}(t) & \propto &t^{z}
\end{array}
\end{equation}

falling within intervals $\delta=0.162\pm 0.020$, $\eta=0.308\pm 0.015$
and $z=1.263\pm 0.040$ from the DP values (we take the DP values of the
dynamical critical exponents from \cite{dp}). On the other hand, the
local slopes approach the asymptotic values, corresponding to the
values of the critical exponents, in a different way depending on the
value of $p_2$.  While for the part of the phase diagram far enough
from the special point (1/2,1) the local slopes stabilize at the
asymptotic value relatively fast (Fig.  \ref{fig:slopes}a), for the
points close to (1/2,1), in the cases we studied, the number of steps
taken was not enough to reach the asymptotic value, although the
approach to DP values was still quite obvious (Fig.
\ref{fig:slopes}b). For $(p_1,p_2)=(1/2,1) $ the exponents agreed with
the theoretical predictions of $\delta=1/2$,$\eta=0$ and $z=1$
\cite{dick}.

The Monte Carlo results for the phase transition line are presented in
Tab.  \ref{tab:tab}. In Fig. \ref{fig:bounds} we compare them with the
rigorous lower bound given by Katori and Tsukahara \cite{kattu} as $p_2=(1-2
p_1^3)/(1-p_1^2)$ and with the rigorous lower ($p_2=2(1-p_1)$) and upper
($p_2=4p_1- 4p_1^2$) bounds given by Liggett \cite{ligg}. Also, we
present the rigorous lower bound obtained by averaging the number of
particles after 12 steps (obtained by exact enumeration, starting from
a single particle) and taking $p_1$ and $p_2$ values leading to the
average equal to 1.
The upper bound given by Liggett comes very
near to the phase transition line in the vicinity of (1/2,1).
It should be remarked that all these bounds can only be proved to be
valid in the so-called attractive region, with $p_1 \leq p_2$ \cite{attr}.
However, we see that the Katori and Tsukahara's lower bound
is not only valid over the whole range of $p_2$ but, indeed,
follows the shape of the Monte Carlo data rather closely.

Fig.3 should be compared with the phase diagram shown by
Kinzel \cite{kinz2}, which was studied by the transfer-matrix scaling
method. Kinzel showed two possibilities for the behavior of the phase
transition line near $p_{2}=0$: the line has an endpoint on the
$p_{1}$-axis at $(p_{1 0},0)$ with $p_{1 0} <1$ with $p_{1 0} <1$, or the line
goes
to the corner point (1,0) in the phase diagram.
The former case is concluded from our results with
$p_{1 0}= 0.8092 \pm 0.0004$.

An interpolation formula, proposed by Yanuka and Englman \cite{yan}
for the ordinary percolation, was shown to perform rather well in the
case of the mixed DP \cite{mixed}. Expanding it to the
general case of the two-neighbor SCA leads to
\begin{equation}
p_2=p_1(2-(p_1/p_1^s)^{1/(1-\lambda)}).
\label{yaneg}
\end{equation}
where $\lambda=\log{p_1^s}/\log{p_2^b}$, with $p_1^s$ and $p_2^b$ as
ordinates of points on the phase diagram, corresponding to the pure
site and pure bond DP, respectively. Although we can not
present any theoretical explanation at this moment, one can see that
(\ref{yaneg}) perform rather well even outside the mixed site-bond
percolation sector, for which it has been derived. Marked discrepancy
begins at $p_2>0.9$, as the phase diagram approaches the special point
(1/2,1).

\section{Series expansion}

While an attempt to directly apply Pade analysis to the series
expansion for the survival probability in $p_1$ and $p_2$ around (0,0)
fails because all coefficients of the expansion change when the order
is increased, changing coordinates to $a$ and $q$ according to
\begin{eqnarray}
p_1&=&1-aq\\
p_2&=&1-q
\end{eqnarray}
leads to a series with converging coefficients. Computer time required
to obtain, using exact enumeration, an expansion to a certain order
grows very fast with the order of the expansion because the model
includes 2 parameters. In the current work we try to take advantage of
the possibility to work analytically with a relatively low order
expansion.

Regarding the series expansion for the logarithm of the survival
probability in $a$ and $q$ as a polynomial in $q$ we take a [3,3] Pade
approximant for its derivative in q and expect that the phase
transition line is given by a root of the denominator (for a
description of the DLog Pade method we are using, for example,
\cite{stauff}). Solving the corresponding 3-rd order equation exactly
and discarding unphysical roots we get a rather complicated expression
for the phase diagram in the form of $q$ given as a function of $a$.
Assuming that we are in the vicinity of the special point, where, as
one can easily see, $a$ goes to infinity, we take an expansion in
$1/a^{1/2}$, the two leading terms are obtained as
\begin{equation}
q=\frac{1}{2a}-\frac{1}{4a^{3/2}}
\label{eq:exp}
\end{equation}
where the discarded terms are of the order $0(a^{5/2})$.
As one can see in Fig \ref{fig:comp}, (\ref{eq:exp}) approximates the
exact solution surprisingly well. Taking (\ref{eq:exp}) and returning to
coordinates $p_1$ and $p_2$ we have for the phase diagram in the vicinity
of (1/2,1)
\begin{equation}
p_2=-3+20 p_1 - 32 p_1^2 + 16 p_1^3.
\label{eq:exa}
\end{equation}
Again, taking a Taylor expansion in $p_1-1/2$ we have
\begin{equation}
p_2=1-8 (p_1-1/2)^2,
\label{eq:tay}
\end{equation}
discarding terms of the order $0((p_1-1/2)^3)$. Surprisingly enough,
comparison with the Monte Carlo data (Fig. \ref{fig:bounds}) shows that
both (\ref{eq:exa}) and (\ref{eq:tay}) work reasonably well in the
vicinity of (1/2,1), with (\ref{eq:exa}), as it should be expected, working
over a wider range of $p_1$ values. Moreover, the parabolic asymptotic
behavior,
predicted by (\ref{eq:tay}), is in agreement with the Ligget's upper
bound and with the Monte Carlo data, as it is shown in Fig.
\ref{fig:par}.

\section{Conclusions and Future Problems}

By using Pade approximations and Monte Carlo simulations, we evaluate the
phase transition line $p_{1}=p_{1 {\rm c}}(p_{2})$
for the two-neighbor SCA. The line has endpoints
at $(p_{1},p_{2})=(1/2,1)$ and at (0.8092,0). Using the time dependent
simulation technique \cite{mixed}, we confirm that the present SCA
belong to the DP universality class in the whole range of parameters,
excluding the point (1/2,1).
The special point (1/2,1) is the transition point of the
reaction-limited case of the dimer-dimer model \cite{takin}
and different critical phenomena are observed \cite{dick}.
The present work show that the shape of the phase transition line
near this special point has, asymptotically, a parabolic shape,
and we conclude that
\begin{equation}
 p_{1 {\rm c}}(p_{2}) - \frac{1}{2} \simeq C
     (1-p_{2})^{\theta}
 \qquad \mbox{with} \quad \theta=\frac{1}{2} \quad \mbox{and}
 \quad C \simeq 0.436
\end{equation}
for $0 < 1-p_{2} \ll 1$.
As discussed in Section 1, this numerical result is a first step
in the further study on the crossover which is expected to
occur in the vicinity of $(p_{1},p_{2})=(1/2,1)$ \cite{durr}.

Recently rigorous lower and upper bounds of the phase transition line
were given for the attractive region $p_{1} \leq p_{2}$
\cite{kattu,ligg}. In this paper we present other rigorous lower bound
obtained by exactly averaging the number of particles after 12 steps,
whose validity can also be proved if and only if $p_{1} \leq p_{2}$.
We access the accuracy of these rigorous bounds using the Monte Carlo
date. The upper bound given by Liggett is excellent
in the vicinity of (1/2,1). It is shown that the lower bound
by Katori and Tsukahara seems to be valid also in the non-attractive region
$p_{1} > p_{2}$ and follows the shape of the Monte Carlo date rather
closely down to $p_{2}=0$.
Justification of the ``lower bound" for $p_{1}>p_{2}$ is a challenging
problem, since rigorous things proved for the non-attractive region are
still limited \cite{KK}.

We also show in this paper that far enough from the special point
(1/2,1) the phase transition line is well approximated by
an interpolation formula \cite{mixed}, similar to the one proposed by
Yanuka and Englman for ordinary percolation \cite{yan}.

\section*{Acknowledgements}

One of the authors (A.T.) acknowledges the support of the Inamori foundation.

\section*{Tables}

\begin{enumerate}
\renewcommand{\thefootnote}{\fnsymbol{footnote}}

\item
Monte Carlo results for the phase diagram.\\
Values marked by \footnotemark[1] were obtained with uncertainty $\pm
0.0002$, \footnotemark[2] corresponds to uncertainty $\pm 0.0004$, and
\footnotemark[3] to $\pm 0.0008$.\\
a) the vicinity of $p_2=1$.\\
b) the vicinity of $p_2=0$.
\label{tab:tab}

\end{enumerate}

\section*{Figure captions}

\begin{enumerate}

\item
The Two-Neighbor Stochastic Cellular Automata
\label{fig:model}

\item
Local slopes for the mean number of particles
$r=\log(n(t)/n(t/5))/\log(5)$ plotted versus the inverse of the time
$t$.\\ Horizontal solid line corresponds to the DP value of the
critical exponent $\eta=0.308$.\\
a) $p_2=0.62$, 1- $p_1=0.9175$, 2- $p_1=0.9171$, 3- $p_1=0.9167$
(critical point), 4- $p_1=0.9163$, 5- $p_1=0.9159$.
\label{fig:slopes}\\
b) $p_2=0.52$, 1- $p_1=0.9983$, 2- $p_1=0.9981$, 3- $p_1=0.9979$
(critical point), 4- $p_1=0.9977$, 5- $p_1=0.9975$.\\
c) $p_1=1/2$, $p_2=1$

\item
Phase diagram for the Two-Neighbor SCA. Monte Carlo results are represented
by points.\\
1,2- lower and upper bounds by Liggett \cite{ligg}.\\
3- lower bound by Katori and Tsukahara \cite{kattu}\\
4- Pade result.\\
5- Tailor expansion for the Pade result.\\
6- Interpolation formula \cite{yan,mixed}.\\
7- lower bound obtained by averaging the number of
particles after 12 steps (obtained by exact enumeration, starting from
a single particle) and taking $p_1$ and $p_2$ values leading to the
average equal to 1.

\label{fig:bounds}

\item
Phase diagram by Pade approximants in coordinates $q$ and $a$,
defined by $p_1=1-q$ and $p_2=1-a q$. The lower line corresponds to
the Taylor expansion, given by
$q=(\frac{1}{2}-\frac{1}{4a^\frac{1}{2}})/a$.
\label{fig:comp}

\item
Phase diagram for the Two-Neighbor SCA in double logarithmic
coordinates. Solid line shows the best fit for $q\equiv
1-p_1=k(p_1-0.5)^2$ over the first 3 points, which corresponds to
$k=5.25\ldots$. Taylor expansions for the Pade result ($k=8$) and
for the Liggett's upper boundary ($k=4$) are also shown.
\label{fig:par}

\end{enumerate}

\newpage
\renewcommand{\thefootnote}{\fnsymbol{footnote}}

\begin{tabular}{ll}
  $p_{1}^c$&   $p_{2}^c$\\
0.50 &   1.0000\\
0.52 &   0.9979\footnotemark[1]\\
0.54 &   0.9917\footnotemark[1]\\
0.56 &   0.9809\footnotemark[1]\\
0.58 &   0.9653\footnotemark[1]\\
0.60 &   0.9443\footnotemark[1]\\
0.62 &   0.9167\footnotemark[2]\\
\end{tabular}

\vspace{4cm}

\begin{tabular}{ll}
  $p_{1}^c$&   $p_{2}^c$\\
0.71 & 0.6877\footnotemark[2]\\
0.73 & 0.6010\footnotemark[2]\\
0.75 & 0.4933\footnotemark[2]\\
0.77 & 0.3590\footnotemark[2]\\
0.79 & 0.1916\footnotemark[3]\\
0.8092\footnotemark[2] & 0.0000\\
\end{tabular}


\begin{thebibliography}{99}
\bibitem{attr}
When $p_{1} \leq p_{2}$, if we add a new particle at a vacant site,
then the probability with which any of the neighboring vacant sites
becomes occupied increases.
It follows that the number of particles at time $t >0$ is a non-decreasing
function of the number of particles at time zero.
This property is called attractiveness.
Using this property, it is proved for $p_{1} \leq p_{2}$
that the survival probability
$P_{\infty}$ is a monotonically non-decreasing function of $p_{1}$
and $p_{2}$
\end{thebibliography}
\end{document}